\documentclass[a4paper,english]{article}

\usepackage{fixltx2e} 
\usepackage[smaller]{acronym}  
\usepackage{cmap} 
\usepackage{ucs}
\usepackage[utf8x]{inputenc}

\usepackage{url}
\usepackage{hyperref}

\usepackage{babel}
\usepackage{enumerate}
\usepackage{ifthen}
\usepackage{graphicx}
\usepackage{relsize}
\usepackage{tabularx}

\usepackage{pslatex}

\usepackage{authblk}
\title{GridCertLib: a Single Sign-on Solution 
  for Grid Web Applications and Portals%
}
\author[1]{R.~Murri}
\affil[1]{%
    Grid Computing Competence Centre\\
    Organisch-Chemisches Institut\\
    University of Z\"urich\\
    Winterthurerstr.~190\\
    CH-8057 Z\"urich\\
    Switzerland 
}
\author[2]{P.~Kunszt}
\affil[2]{%
    SystemsX, ETH Z\"urich\\
    Clausiusstr.~45\\
    CH-8092 Z\"urich\\
    Switzerland
}
\author[1]{S.~Maffioletti}
\author[3]{V.~Tschopp}
\affil[3]{%
    SWITCH\\
    Werdstrasse 2\\
    CH-8004 Z\"urich\\
    Switzerland
}
\date{Sep.~9, 2011}

\providecommand*{\DUprovidelength}[2]{
  \ifthenelse{\isundefined{#1}}{\newlength{#1}\setlength{#1}{#2}}{}
}

\DUprovidelength{\DUdocinfowidth}{0.9\textwidth}

\ifthenelse{\isundefined{\hypersetup}}{
  \usepackage[colorlinks=true,linkcolor=blue,urlcolor=blue]{hyperref}
}{}
\hypersetup{
  pdftitle={GridCertLib: a Single Sign-on Solution for Grid Web Portals},
  pdfauthor={Riccardo Murri (GC3 / University of Zurich);Peter Kunszt (SystemsX.ch / ETH Zurich);Sergio Maffioletti (GC3 / University of Zurich);Valery Tschopp (SWITCH)}
 }

\begin{document}

\maketitle

\acused{API}%
\acused{GAMESS}%
\acused{HTTP}%
\acused{SARoNGS}%
\acused{SWITCH}%
\acused{URL}%


\section*{Abstract%
  \phantomsection%
  \addcontentsline{toc}{section}{Abstract}%
  \label{abstract}%
}


This paper describes the design and implementation of
\emph{GridCertLib}, a Java library leveraging a Shibboleth-based
authentication infrastructure and the SLCS online certificate signing
service, to provide short-lived X.509 certificates and Grid proxies.

The main use case envisioned for \emph{GridCertLib}, is to provide
seamless and secure access to Grid X.509 certificates and proxies in
web applications and portals: when a user logs in to the portal using
SAML-based Shibboleth authentication, \emph{GridCertLib} uses the SAML
assertion to obtain a Grid X.509 certificate from the SLCS service and
generate a VOMS proxy from it.

We give an overview of the architecture of \emph{GridCertLib} and
briefly describe its programming model.  Its application to some
deployment scenarios is outlined, as well as a report on practical
experience integrating \emph{GridCertLib} into portals for
Bioinformatics and Computational Chemistry applications, based on the
popular P-GRADE and Django softwares.


\section{Introduction%
  \label{introduction}%
}

Most Grid computing middleware in production use today relies on X.509
certificate proxies \cite{RFC3820} for user authentication.  This has
been an issue when implementing web-based interfaces to Grid computing
facilities: in order to generate a proxy, a copy of the X.509 private
key is needed together with the passphrase used to encrypt
it. However, uploading the public/private key pair to a web portal is
undesirable on security grounds.  Several solutions and workarounds
have been implemented (see
Section~\ref{an-overview-of-existing-solutions} below), but none of
them can be considered entirely satisfactory: either because they do
not fully address the security concerns, or because they require end
users to take multiple steps, possibly through different and unrelated
user interfaces (e.g. a web portal and UNIX shell commands).

The solution we developed leverages two features offered by SWITCH:
the federated authentication and authorization infrastructure
SWITCHaai and the \ac{SLCS}. SWITCHaai is a
federated authentication and authorization infrastructure
\cite{SWITCHAAI}, based on Shibboleth2 \cite{SHIB2}; SWITCHaai
federates all Universities in Switzerland, plus major research centers
and educational institutions.  Similar nationwide \acfp{AAI} exist in
Germany, Denmark, Belgium, the Netherlands and other countries
\cite{Federation}.  A web service provider (e.g., a portal) requiring
SWITCHaai authentication will delegate the authentication step to the
user's home institution \acf{IdP}.  Users will be prompted with the
familiar login page of the home institution; after successful logon,
the service provider will receive a set of parameters and additional
metadata (about the user) to proceed with authorization
\cite{SWITCHaaiDemo}.

SWITCH also provides the \acf{SLCS} and makes it available to all
SWITCHaai users \cite{SLCS}. \ac{SLCS} is a web-service that can sign
an X.509 certificate online; authentication and authorization to the
\ac{SLCS} service are based on the SWITCHaai Shibboleth system.  The
online \acf{CA} that signs \ac{SLCS} certificates is included in the
\ac{IGTF} bundle \cite{IGTF}, so \ac{SLCS} certificates can be used
for any legitimate Grid purpose.  This enables any user from a Swiss
institution participating in the SWITCHaai Federation to request a
Grid-enabled X.509 certificate. It is valid up to 1'000'000 seconds
(corresponding to almost 11 days) which is short-lived in comparison
to regular one-year certificates issued by other CAs. A \ac{SLCS}
command-line client is also part of the gLite middleware distribution
\cite{gLite-SLCS}.

The last key enabler for our project is the Shibboleth delegation
feature, developed in the Shibboleth uPortal project
\cite{ShibUPortal}. The delegation is based on the Liberty
\acs{ID-WSF} \acs{ECP} \ac{SSO} profile \cite{ECPProfile}, and allows
SAML-based authentication for Shibboleth-protected web service
providers.\footnote{%
  The interested reader can find readable introductions to Shibboleth
  and \acs{SAML} in \cite{SHIB2intro,SAMLintro}.
}%
Based on the assertion resulting from the web authentication through
Shibboleth on the user portal, we are able to call the \ac{SLCS}
service on the user's behalf using the \ac{ECP} profile. Delegation,
however, is still an experimental feature in Shibboleth, and is
expected to become standard in the next Shibboleth version 3.0.  For
our project, SWITCH has upgraded their Shibboleth Virtual Home
Organization identity provider and the \ac{SLCS} service provider with
\ac{ECP} delegation features.

\emph{GridCertLib} is a Java library providing programming interfaces
to create a \ac{SLCS} certificate and Grid proxy (optionally
\acs{VOMS}-enabled), given the \ac{SAML} assertion resulting from a
successful previous Shibboleth authentication.  The main use case
envisioned for \emph{GridCertLib}, is to provide seamless and secure
access to Grid X.509 certificates and proxies in web portals: when a
user logs in to the portal using the regular SWITCHaai Shibboleth
authentication, \emph{GridCertLib} uses the \ac{SAML} assertion to
obtain a Grid X.509 certificate from the \ac{SLCS} service and
generate a \ac{VOMS} proxy from it. None of these steps requires user
interaction (after the initial Shibboleth authentication), making
Grid resources as easy to use as any single-sign-on enabled web
service while retaining the full security stack.

The outline of the paper is as follows: first we provide an overview of
similar solutions already implemented in production-grade Grid web
portals.  In the next Section, we review the requirements that were
set for \emph{GridCertLib}, its actual design and discuss some
implementation details.  
Finally, we report on some deployment scenarios and particularly on
the integration of \emph{GridCertLib} within a Bionformatics portal
(based on P-GRADE, \cite{PGRADE}) and within a Computational Chemistry
portal (based on Django, \cite{Django}).


\section{An overview of existing solutions%
  \label{an-overview-of-existing-solutions}%
}

Distributed authentication and authorization is a difficult problem to
solve in a standard and integrated manner. In past Grid projects,
proprietary services have often been developed to address this issue
(e.g. CAS\cite{cas}, PRIMA\cite{prima}, ROAM\cite{roam}) but none of
them established itself as a widely accepted community standard as
they were too tightly coupled with the middleware and the local
resources. A standardization on \ac{SAML}/\acs{XACML} profiles to be used by all
middlewares is available \cite{samlprofile} but has not been widely
adopted yet.

However, we can see two technologies that are widely accepted also
outside of the Grid community: \ac{SAML}-based authentication using
Shibboleth and X.509 certificates to authenticate local resources,
using short-lived proxy certificates. In most Grids, also the
\ac{VOMS} cervice is used to enrich the proxy certificate with usage
attributes for fine-grained authorization. We are using the \ac{SLCS}
service as developed by SWITCH for the \ac{EGEE} consortium to generate
certificates from the users' Shibboleth login. A similar but now
defunct project was the U.S. GridShib effort \cite{GridShib} to create
a certificate based on a Shibboleth login also as a Certificate
Authority. 

In order to generate a certificate proxy, a copy of the X.509 private
key is needed together with the passphrase used to encrypt it.  This
poses a basic problem in web portals: having direct access to the
public/private certificate key pair of a user, although technically
feasible, is undesirable on security grounds: intruders getting access
to the portal machine would gain unrestricted access to all of the
portal users' credentials.

Some projects have worked around this issue by submitting to the Grid as a
single portal superuser, using credentials of a single entity for all
Grid jobs issued through the portal or through special-purpose
certificates for automation, called ``robot'' certificates.

Robot certificates are X.509 certificates granted to a portal service
or application, rather than a human; users interested in running a
certain application on the Grid can log in to the portal, and the
portal will operate on the Grid using the robot certificate.  This
approach a few drawbacks:
\newcounter{listcnt0}
\begin{list}{\arabic{listcnt0}.}
{
\usecounter{listcnt0}
\setlength{\rightmargin}{\leftmargin}
}

\item The certificate private key is available on the portal machine,
  although this can be prevented by using hardware-based protection
  (e.g. smartcards), as done in the GILDA/GENIUS portal
  \cite{RobotGilda}.  Indeed, guidelines \cite{RobotCertificates} have
  been issued by the \ac{IGTF} on the generation and storage of private
  keys, and permissible key usage of automated clients (robots) that
  can hold credentials issued by \ac{IGTF}-accredited Certification Authorities, so this
  specific issue is likely to become less relevant in the future.

\item The use of robot certificates moves the responsibility of user
  authentication and logging from the \ac{CA} to the portal, thus
  implicitly introducing an additional trust step in the Grid
  authentication infrastructure.  Not all Grid sites and resource
  providers might be happy with delegating trust this way.

\item It is difficult to provide per-user accounting of computational
  resource usage: jobs submitted through different interfaces (e.g.,
  portal and command-line) by the same user will be accounted to
  different end-entities, since all popular Grid middlewares group
  usage records by certificate subject \ac{DN}.
\end{list}

The solution adopted in the P-GRADE portal \cite{PGRADE,PGRADE2} is to
have users upload a long-lived proxy to a MyProxy server
\cite{MyProxyHomePage,MyProxy1,MyProxy2} and authorize
the portal software for automated retrieval of short-lived proxies for
job submission and data movement.  However, this still requires users
to deal with many of the complexities of managing X.509-based
certificates and command-line tools, which has been found to be a real
barrier to Grid adoption in less tech-savvy user communities.  In the
newer WS-PGRADE portal \cite{wspgrade} the interaction with the
MyProxy service has been streamlined so no command-line interaction is
necessary, user certificate can be directly uploaded. However, the
user still needs to apply for and manage a certificate that expires
every year. For the end-user it is a complication to use an
authentication infrastructure that does not blend with the native web
portal authentication system. It interrupts the natural flow of
operations in the web user interface, requiring either an additional
password (the certificate password to generate the proxy) or
additional command-line operations in order to proceed with Grid job
submission and control.

An extension to this mechanism that blends more seamlessly with
P-GRADE's web-based interface has been developed by the UK project
SARoNGS in \cite{SARoNGS-PGRADE}.  Clicking a button on the MyProxy
web details page redirects the user to a web service
(Shibboleth-protected), which in turn loads a long-lived proxy into a
specific MyProxy server, and fills in the details in the P-GRADE
configuration page.

The approach taken in \emph{GridCertLib}, instead, requires no user
interaction: once the web-based Shibboleth login is successfully
completed, the \emph{GridCertLib} code can generate an X.509 certificate
through the \ac{SLCS} service using the web service based \ac{ECP} delegation,
and an accompanying Grid proxy.  Details of this process are given in
the following sections.

The source code of \emph{GridCertLib} is publicly available from
\href{http://gridcertlib.googlecode.com/}{\url{http://gridcertlib.googlecode.com/}} under the Apache License 2.0
\cite{ApacheLicence2}.


\section{Design and Implementation%
  \label{design-and-implementation}%
}


\subsection{Architecture overview%
  \label{architecture-overview}%
}

\emph{GridCertLib} was designed to bridge Shibboleth-based and Grid
X.509-based authentication services for web applications and
portals.\footnote{Henceforth, we shall briefly write ``portal'' to mean
  any web-based interactive application or service.}  Its design goals
were to allow easy integration into any Java portal, and to minimize
interaction with the user while retaining the full security stack for
Grid authentication and authorization.
\begin{figure}
  \begin{center}
    \includegraphics[bb=0 0 612 612,width=1.0\textwidth]{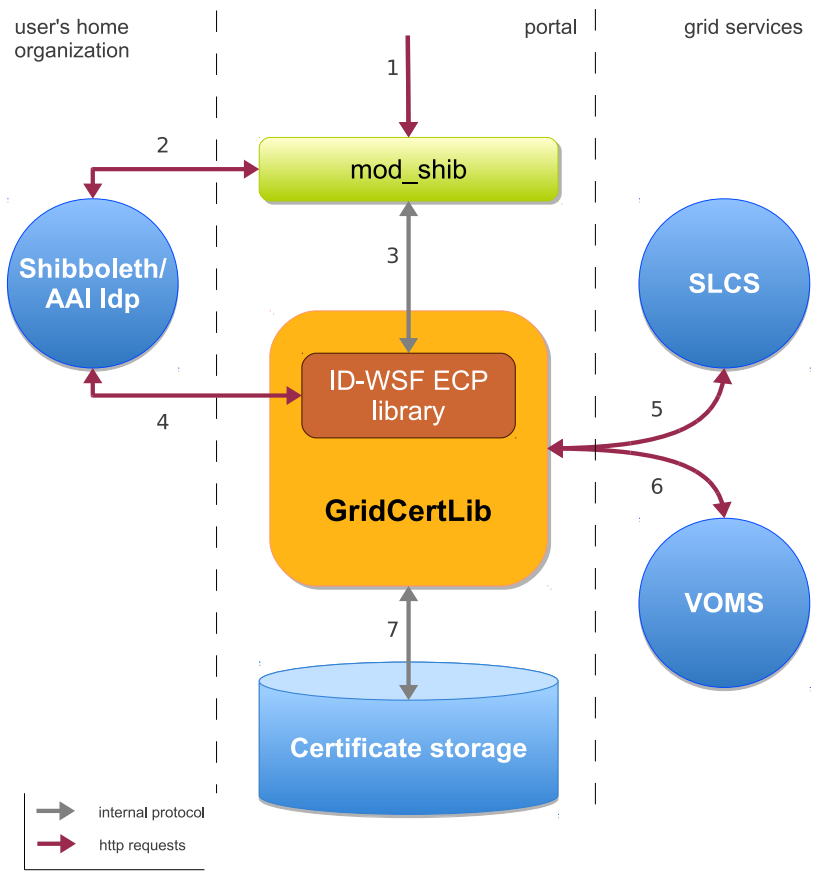}
    \caption{Interactions of \emph{GridCertLib} with the other
      involved services.  Components in the center column are all part
      of the same web application (but they could be part of different
      processes: for instance, the Shibboleth web login is usually
      handled by an Apache module rather than by Java servlet code).
      Round boxes on the sides represent services running on remote
      hosts, that \emph{GridCertLib} interacts with over \acsu{HTTP}.  A
      detailed description of interactions is given in
      Section~\ref{architecture-overview}.}
    \label{fig:interactions}
  \end{center}
\end{figure}

The flow of interaction with the Java portal code and the SWITCHaai
services illustrated in Figure 1 was devised in order to accomplish
the design objectives (numbers in parentheses correspond to arrows in
Figure~\ref{fig:interactions}):
\begin{enumerate}[\em (1)]

\item Users initiates log in to the web portal using Shibboleth
  single sign-on (i.e., request a Shibboleth-protected URL).

\item They are authenticated by their home organization's identity
  provider \ac{IdP}; this is handled transparently by the
  Shibboleth software.\footnote{%
    A detailed understanding of the Shibboleth authentication process
    is not needed for working with \emph{GridCertLib}: it suffices to
    know that the outcome of a successful Shibboleth logon is a
    \ac{SAML2} assertion stored in the web server on the Portal
    machine.  The interested reader is referred to
    \cite{SHIB2intro,SWITCHaaiDemo,SAMLintro} for an introduction to
    Shibboleth and \ac{SAML2}.
    }%

\item \emph{GridCertLib} queries Apache's \texttt{mod\_shib} to get
  the \ac{SAML2} assertion. The assertion is exported, together with
  other authentication parameters, to any proxied web service; portals
  may make use of these Shibboleth attributes to restrict certain
  services or map users into user groups.

\item The portal application code calls GridCertLib to obtain a
  short-lived Grid X.509 certificate signed by the \ac{SLCS} \ac{CA}.
  This step requires delegation of the Shibboleth credentials
  (\ac{SAML2} assertion) to the \ac{SLCS} login service, which is done
  through the generic \ac{ID-WSF} \ac{ECP} Web Service Client,
  developed by SWITCH \cite{ECP}.

\item After logging in to the \ac{SLCS} service, \emph{GridCertLib}
  proceeds to generate an X.509 certificate and have it signed by the
  \ac{SLCS} online \ac{CA}, using code similar to the one used by
  gLite's \texttt{slcs-init} command-line client \cite{gLite-SLCS}.

\item The portal calls GridCertLib to create a grid proxy (with or
  without \ac{VOMS} extensions).  Here \emph{GridCertLib} is just a
  thin wrapper around the regular \ac{VOMS} libraries, mainly
  providing simpler façade calls for commonly-used cases.

\item The certificate, private key and proxies are stored.
  Currently, \emph{GridCertLib} provides methods for persisting
  certificates and proxies to the filesystem; it is up to
  the portal to move the files to different storage back-ends (e.g.,
  databases), although it should be noted that most Grid middlewares
  require the proxy to be in a known location on the filesystem.
\end{enumerate}

As soon as the \emph{GridCertLib} code completes successfully, a valid
certificate and proxy are available on the filesystem and Grid
operations can proceed.

The foreseen usage of \emph{GridCertLib} in web portals (see Section
\hyperref[deployment-experiences]{Deployment Experiences}) called for
implementation of additional features in \emph{GridCertLib}:
\setcounter{listcnt0}{0}
\begin{list}{{\em \alph{listcnt0}.}}
{
\usecounter{listcnt0}
\setlength{\rightmargin}{\leftmargin}
}

\item The \ac{SLCS} and the Grid/VOMS proxy generation functions can be called
independently. In particular, proxy generation does not rely on the
\ac{SLCS} generation feature being called first.  As a consequence, Grid
proxy code does not need Shibboleth authentication to run, it only
expects a valid user certificate to be available.

This is a portal user interface requirement: all that is necessary
for generating an X.509 certificate is known after \ac{SLCS} login, but
proxy generation requires additional data; namely, names of the VOs
a user belongs to.  Portals might need to gather this additional
data \emph{after} a user has logged in.

\item The \ac{SLCS} generation function can be called at any time, as long as
the \ac{SAML2} assertion resulting from the Shibboleth login process is
available.

This is another portal user interface requirement: \ac{SLCS} login and
generation of X.509 certificates can take long times (from a user interface
perspective), so they can be delayed at a later stage or started in
an asynchronous thread, in order not to delay the login process.

\item \emph{GridCertLib} has a generic interface that can be used with any web
application or portal.  In particular, \emph{GridCertLib} does not make
any assumption on how user data (like the certificates) are
represented and/or stored within the portal, except that they can
be stored on the filesystem.
\end{list}

Feature \emph{a.}\ led us to provide \emph{GridCertLib} with two main independent
entry points, \emph{SLCSFactory} and \emph{GridProxyFactory}.  An instance of
each class is responsible for generating \ac{SLCS} certificate and Grid
proxies, respectively.  To achieve portal independence, each class constructor takes
an explicit list of all the parameters needed for instantiation,
although they can also be conveniently provided by a single Java
\emph{Properties} object.

Similarly, for the same goal \emph{c.}\ of portal technology independence,
\emph{GridCertLib} certificate and proxy generation functions accept an
explicit list of all the required parameters, but provide shortened
forms that have common defaults.

Feature \emph{b.}\ implies that the \acs{SLCS}-generation methods in
\emph{GridCertLib} only require the Shibboleth \acs{SAML} assertion as
input.  However, the \ac{SAML2} assertion can expire long before the
Shibboleth session itself does (see Section~\ref{creating-slcs-certificates}).
To solve this problem, \emph{GridCertLib} provides a re-usable servlet
\hyperref[renewassertion]{RenewAssertion}, which can also be used as a
model for implementing assertion renewal in the portal code.


\subsection{Core Library Implementation%
  \label{core-library-implementation}%
}

\emph{GridCertLib} core functions reside in the single Java package
\texttt{ch.swing.gridcertlib}; an additional package
\texttt{ch.swing.gridcertlib.servlet} provides example servlets (with
fully-commented code) that show how the library can be used.

The main package \texttt{ch.swing.gridcertlib} has two public entry
points:
\begin{itemize}

\item The \emph{SLCSFactory} class provides the \ac{SLCS} certificate generation
functionality and can store the certificate and its associated
private key on the filesystem.

\item The \emph{GridProxyFactory} class creates Globus Toolkit proxy
certificates with or without \ac{VOMS} extensions from available user
certificates and stores them to a temporary location on the
filesystem.

\end{itemize}

A single instance of each of these classes can generate multiple \ac{SLCS}
certificates or proxies, possibly for different Portal users, via
repeated invocation of the certificate creation methods.

Since the parameters used to configure the factory objects are
portal-wide global variables and their values are fixed while the
portal application is running, each factory class can be configured
(at construction time) through a \emph{java.{\hspace{0pt}}util.{\hspace{0pt}}Properties}
object, which can be conveniently loaded from a file with standard
Java \ac{API} calls.  Alternatively, a constructor that allows to specify
all instance parameters explicitly is also provided.

However, \emph{GridCertLib} does not enforce that only a single instance of
these factory objects exists. Different factory objects can be
created to cater for different classes of users (e.g., users coming
from different Shibboleth federations).  It is up to the web
application/portal code to route requests to the correct factory.


\subsubsection{Creating SLCS Certificates%
  \label{creating-slcs-certificates}%
}

Upon calling the \emph{newSLCS} method of the \emph{SLCSFactory} class, a
\emph{SLCSRequestor} object is created to carry out generation of the
certificate and actual interaction with the \ac{SLCS} server.  The reason
for this split is twofold:
\begin{itemize}

\item \emph{SLCSFactory} handles system-wide defaults, and thus a single
instance is needed to serve the whole portal, whereas a new
\emph{SLCSRequestor} object is created for every certificate request.

\item The \emph{SLCSRequestor} corresponds to the \texttt{slcs-init} command-line
tool provided in the gLite middleware distribution \cite{gLite-SLCS};
this eases porting of fixes from the official \ac{SLCS} client on to
\emph{GridCertLib}.

\end{itemize}

\ac{SLCS} certificates are created following these steps:
\setcounter{listcnt0}{0}
\begin{list}{\arabic{listcnt0}.}
{
\usecounter{listcnt0}
\setlength{\rightmargin}{\leftmargin}
}

\item Login to the \ac{SLCS} server using \ac{ECP} delegation: if successful, this
returns the subject \ac{DN} to use in the X.509 certificate to be
generated, and an authorization token to validate the final
certificate signing request to the same \ac{SLCS} server.

\item Locally generate an X.509 public/private key pair.

\item Locally generate an X.509 \acf{CSR},
using the subject \ac{DN} and other X.509 constraints returned by step 1.

\item Submit the \ac{CSR} to the \ac{SLCS} server and get the signed certificate back.
\end{list}

All of the above steps keep data (like the password), which is necessary
for the private key, in memory.  Only after a certificate has been
successfully generated by the \emph{SLCSRequestor} will \emph{SLCSFactory} save
the result in a file and return the certificate file path, private key
path and private key password to the caller. Any of these three can be
defined by the client code by passing an optional argument to the
newSLCS method; by default, \emph{SLCSFactory} uses a random password and
stores the certificate and private key files in a configurable
directory (using a random file name, which is also returned as a result of
the call).

The \ac{SLCS} service is a properly authorized Shibboleth \acf{SP}. Since \emph{GridCertLib} is contacting \ac{SLCS} on behalf of the user,
and with no user intervention at all, delegation of \ac{SAML} credentials
is needed.  Shibboleth delegation is an experimental feature available
in Shibboleth2, by which the \ac{SAML2} assertion that initiates a
Shibboleth session on an \ac{SP}, may be re-used to authenticate towards
other web service based SPs.  Shibboleth delegation must be supported
both on the \ac{IdP} granting the \ac{SAML2} assertion \emph{and} the target \ac{SP}
receiving the delegated assertion (the \ac{SLCS} server in this case).

The requirement that \ac{SLCS} generation may happen at any time after
login to the portal creates an additional complication: \ac{SAML2}
assertions have a short time validity (5 minutes in the default
configuration) but \ac{SP} and \ac{IdP} sessions last much longer (8 hours by
default), i.e., users are authenticated with the Shibboleth \ac{SP} even
after the \ac{SAML2} assertion is long expired.  Therefore, by the time
\texttt{SLCSFactory.newSLCS} is called, the \ac{SAML2} assertion might be
unusable for delegated authentication with the \ac{SLCS} server.  There is
no support in the Shibboleth \ac{API} to get a fresh assertion in the \ac{IdP}
and store it in the \ac{SP} session, but this can be worked around by
forcing an \ac{SP} session logout, followed by a redirection to a
Shibboleth-protected \acs{URL}: the \ac{SP} will start a new session and request
a fresh assertion from the \ac{IdP}.  This can be implemented by a chain of
\ac{HTTP} redirections, so that the whole procedure does not require any
user intervention.  We implemented this workaround in the
\texttt{RenewAssertion} servlet, described in
Section~\ref{example-servlets}.


\subsubsection{Creating Grid and VOMS proxies%
  \label{creating-grid-and-voms-proxies}%
}

The \emph{GridProxyFactory} class is a interface wrapper on top of the \ac{VOMS}
Java \ac{API}.  \emph{GridProxyFactory} implements a simplified interface to
create a Grid proxy in the use case most frequently needed in web
applications and portals: its \emph{newProxy} method creates a proxy (with
optional \ac{VOMS} extensions) given an X.509 certificate and private key,
and a (possibly empty) list of VOs to contact for \ac{VOMS} ACs.

A single instance of the class can generate multiple proxies (possibly
for different users) via repeated invocation of the \emph{newProxy}
method.  Since the \emph{org.{\hspace{0pt}}glite.{\hspace{0pt}}voms.{\hspace{0pt}}contact.{\hspace{0pt}}VOMSProxyInit}
class uses system properties to determine part of its configuration,
it is not possible to create different instances of this class, each
using its own configuration.  This is not a limit in practice, as the
\emph{org.glite.voms} library has native support for multiple servers
and \ac{VO} endpoints.


\subsubsection{Example servlets%
  \label{example-servlets}%
}

The provided sample servlets can run in any Java servlet container.
They have been successfully tested with the Jetty and Tomcat Java
application servers (with an Apache proxy front-end for managing the
Shibboleth session).

Three servlets are distributed with the GridCertLib source code:
\begin{itemize}
\item \emph{SlcsInit:} This servlet generates an X.509 certificate and
  private key and uses the \ac{SLCS} service to sign it.  Upon
  successful completion, the certificate and private key are stored in
  the filesystem.
\item \emph{VomsProxyInit:} This servlet creates a \ac{VOMS} proxy and
  stores it in the filesystem.
\item \emph{RenewAssertion:} This servlet ensures that a fresh
  \ac{SAML} assertion is stored in the \ac{SP} Shibboleth session
  cache.
\end{itemize}
A detailed description of each of these servlets follows.


\paragraph{SlcsInit.%
  \label{slcsinit}%
}

The \emph{ch.swing.gridcertlib.servlet.SlcsInit} servlet extracts the
\ac{SAML2} assertion \ac{URL} from the Shibboleth \ac{HTTP} headers, downloads the
assertion into memory, and uses it to authenticate to a remote \ac{SLCS}
service and get a new certificate/private key pair.  The key is
encrypted with a random password, and the certificate and private key
locations (on the filesystem) are printed in the response text.

If \emph{SLCSFactory} detects an expired assertion in the \ac{SP} session, it
will raise an exception. The \emph{SlcsInit} code catches the error and
redirects the user's browser to the \hyperref[renewassertion]{RenewAssertion} servlet, setting
the return address to the current page: when the user browser is sent
back to the return \ac{URL}, a new \ac{SAML2} assertion will be in the \ac{SP} cache.


\paragraph{VomsProxyInit.%
  \label{vomsproxyinit}%
}

The \emph{ch.swing.gridcertlib.servlet.VomsProxyInit} servlet creates a
\ac{VOMS} proxy and stores it on the filesystem, in the default store
directory.  \ac{HTTP} query parameters can set arguments that are passed to
the \emph{GridProxyFactory.newProxy} method, thus making this servlet a
generic front-end to the \emph{GridProxyFactory} class functionality.

This servlet does not require any interaction with the Shibboleth
subsystem, and can be deployed unprotected.  It requires, however,
that the certificate and private key are available on the
filesystem.


\paragraph{RenewAssertion.%
  \label{renewassertion}%
}

The \emph{ch.swing.gridcertlib.servlet.RenewAssertion} servlet ensures
that a fresh assertion is stored in the \ac{SP} Shibboleth session
cache.  It implements the workaround described in a previous section
for the ``expired assertion'' problem: \setcounter{listcnt0}{0}
\begin{list}{\arabic{listcnt0}.}
{
\usecounter{listcnt0}
\setlength{\rightmargin}{\leftmargin}
}

\item The user's browser is redirected to the \ac{SP} session logout \ac{URL}.

\item The logout function allows setting a ``return address'' via a \ac{URL}
query parameter, to which the browser will be redirected after the
logout is done; this ``return address'' is set to the
\emph{RenewAssertion} \ac{URL} plus a trailing component (\ac{URL} ``path
information'' part) that encodes the referring page \ac{URL}.

\item The Shibboleth \ac{SP} logs the user out of the session and destroys the
cached data, then redirects the user browser to the
\emph{RenewAssertion} \ac{URL}.

\item The \emph{RenewAssertion} page is Shibboleth-protected, so a new
Shibboleth authentication procedure begins.  As long as the user
session in the \ac{IdP} is still valid, this will not require user
interaction, and the \ac{IdP} will just send a new \ac{SAML2} assertion to
the requesting \ac{SP}.

\item The \emph{RenewAssertion} servlet detects that the browser is returning
after the initial visit (from the trailing portion of the \ac{URL}), and
redirects the user to the initial requesting page (by decoding the
\ac{URL} embedded in the ``path information'' component).
\end{list}

Note that none of the above steps requires any user interaction
(unless the Shibboleth session on the \ac{IdP} is expired).

A request \ac{URL} to RenewAssertion must be properly formatted; the
convenience method \emph{RenewAssertion.getRenewalUrl} is provided to this
purpose.  However, the \ac{URL} encoding system in the \emph{RenewAssertion}
servlet imposes a limit on the length of return URLs; more
importantly, it cannot be used with \ac{HTTP} POST requests, as there is no
way of encoding the POST data into a single \ac{URL}.  This is a technical
issue which we have not been able to work around so far: due to the
large number of \ac{HTTP} redirects taking place, session cookies, query
parameters, and other commonly-used ways of associating state data
with \ac{HTTP} requests, may be lost before the final visit to the
\emph{RenewAssertion} servlet.


\section{Deployment experiences%
  \label{deployment-experiences}%
}

The following points need to be taken into consideration by the portal
providers:
\begin{itemize}

\item Since certificate generation can be time-consuming (relative to user
interface reaction times), it could be delayed to a later stage or
executed asynchronously in a separate thread. However, this delay was not
a problem on the P-GRADE Bioinformatics portal.

\item The validity of the Shibboleth assertion is usually limited to a few
minutes, so the \ac{SLCS} certificate request should not be delayed for
too long. Of course if a valid \ac{SLCS} certificate for the user is
already available from a previous login of the user, the request can
simply be omitted.

\item When a \acs{VOMS}-enabled proxy is needed, it is the portal's
responsibility to prompt the user for the relevant information, e.g., \ac{VO}
name or \ac{FQAN} list. In the P-GRADE implementation, the users can set
their VOs in their settings page. There is a global default
configuration option for the administrator if every user is expected
to be always member of the same \ac{VO}.

\end{itemize}


\subsection{Integration into the P-GRADE portal%
  \label{integration-into-the-p-grade-portal}%
}

The P-GRADE portal comes with full Grid X.509 proxy support, which in this case is a mixed
blessing as many of the certificate management features need to be modified in various places of the
portal code. Out of the box, P-GRADE supports proxy certificate upload or the usage of a MyProxy
server to which the user has to upload the certificate outside of P-GRADE.

In its standard form, P-GRADE provides no facilities for the creation
of the certificates; this is a new feature we add using
\emph{GridCertLib}. We extended the Shibboleth-enabled login
\cite{GridSphereAndShibboleth} for the Gridsphere portal \cite{GridSphere}
(provided by the Australian MAMS project \cite{MAMS}) by storing all
Shibboleth attributes including the assertion and other attributes
that were not previously requested into a singleton object.

In the MAMS implementation, on first-time login using Shibboleth, the
user is presented with a registration request portlet which simply
displays the attributes of the user as received through the Shibboleth
login by the server. Users can then simply press a button ``Send
registration request'', which triggers an email to the portal
administrator, who can decide whether to enable the user account, and
optionally assign it certain roles in Gridsphere.

Users can simply reload the page or re-login once the admin has
enabled them. At the same time it is checked whether an \ac{SLCS}
certificate still exists for the given user and whether it is valid
for longer than 24 hours. If not, \emph{GridCertLib} is used to create
a new \ac{SLCS} certificate. The certificate location and other
related information is stored together with all other user attributes
in the user table, which has been extended accordingly.

The \ac{VOMS} configuration is the same for all users of the portal in our
current implementation, which is set to the ``life'' \ac{VO} of the national
grid computing infrastructure \ac{SMSCG} \cite{SMSCG}, using a portal-wide
configuration of \emph{GridCertLib}.

An issue remains: the delegation feature used by \emph{GridCertLib} is not yet
deployed as a standard feature in the Swiss SWITCHaai federation,
therefore we currently can only make use of this whole mechanism
through a special home organization, the \acf{VHO}, provided by
SWITCH for collaboration purposes. We have a dedicated group in the
\ac{VHO} where we can administer our own users. This should not be
necessary anymore after the SWITCHaai federation has upgraded to a
version of Shibboleth that supports delegation, which should happen
sometime in late 2011 or 2012.

For now, in the optimal case a user can log in through \ac{AAI} by
selecting the \ac{VHO} as the ``home organization'', and is ready to submit
Grid jobs to the Swiss Multi-Science Computing Grid. Clicking on the
``Certificates'' tab will show the details of the current certificates
and their validity.

Expiration of the certificate is not an issue, as P-GRADE requests the
download of the results only when the user asks for it through the
portal. The portal makes sure that a new proxy is generated
automatically in the background from the \ac{SLCS} certificate (if the
existing proxy is not valid anymore).  Should the \ac{SLCS}
certificate expire, a new one is requested automatically at the next
login, so unless the user is actively using the portal browser window
for 10 days with no interruption, this will not happen.


\subsection{Integration into Django-based web applications}
\label{sec:django}

Django \cite{Django} is a high-level Python Web framework, providing
reusable components to build any sort of web application.  We have
used it to build a simple portal for users of the computational
chemistry application GAMESS-US \cite{GAMESS:1993,GAMESS:2005}.  
The portal uses the \emph{django-shibboleth}
application\footnote{Django structures a web site as an set of web
  applications, each of which is attached to specific URLs in the web
  site \ac{URL} space.  Applications can be packaged and deployed
  separately, and can be thus re-used in different combinations to
  build a site.} \cite{DjangoShibboleth} to enable users to log in
using their SWITCHaai/Shibboleth credentials; new users will have
their account created automatically when they log in for the first
time.

Django support in \emph{GridCertLib} thus comprises two
(inter-dependent) parts:
\begin{itemize}
\item A Python package, containing the access-control
  decorators\footnote{Django routes \ac{HTTP} requests to Python functions
    (``view'' functions), that are responsible for returning content
    to the user.  Access control is most easily done through Python
    function decorators: if a view function is marked with the
    \emph{login\_required} decorator, then Django ensures that \ac{HTTP}
    requests to that \ac{URL} come from logged-in users, and will redirect
    any unauthorized request to the site login page.}
  \emph{certificate\_required} and \emph{gridproxy\_required}.  By
  using these decorators, a Django programmer can easily mark some
  URLs as requiring the use of a valid \ac{SLCS} certificate and/or proxy.
\item A set of Java servlets, which should be deployed alongside the
  Django site, that interface with \emph{GridCertLib} to provide the
  SLCS- and proxy-generation functionality.
\end{itemize}
All communication between the Django decorators and the corresponding
servlets happens by means of \ac{HTTP} redirects through the users' web
browser.

The \emph{GridCertLib} Django decorators will first ensure that the
\ac{HTTP} request is authenticated with the standard Django login system;
when the \emph{django-shibboleth} application is installed, this
automatically ensures that the \ac{HTTP} request is part of a valid
Shibboleth session.

Next, \emph{GridCertLib} Django decorators check that the certificate
(resp.\ proxy certificate) exists and is valid.  For the sake of
processing speed (no response can be sent to the web browser until the
decorator has passed control to the view function), the decorators
assume that no other actor can modify the certificate/proxy files they
have created: thus a simple ``modification time'' check suffices to
prove that a certificate/proxy is still in its validity period.
Note that, in contrast to what happens in the P-GRADE portal, the
certificate/proxy check happens each time the Django view function is
invoked, and it is thus essential to keep it performant.

If the certificate/proxy exists and is valid, environment variables
are set to the filesystem path of the relevant files to communicate
the location to the Grid middleware, and control is passed to the view
function.

Otherwise, an \ac{HTTP} redirect response is issued, channeling the
web browser to the \ac{URL} corresponding to a Django-specific version
\hyperref[slcsinit]{\em SlcsInit} or
\hyperref[vomsproxyinit]{\em VomsProxyInit} servlets.  As mentioned for
the Example Servlets (see Section~\ref{vomsproxyinit}), only the
\mbox{\hyperref[slcsinit]{\em SlcsInit}} URL needs to be Shibboleth-protected.

\ac{HTTP} session cookies are used to tell the servlets to store the
certificate/proxy in a certain filesystem location;\footnote{Since the
  \hyperref[slcsinit]{\em SlcsInit} and
  \hyperref[vomsproxyinit]{\em VomsProxyInit} servlets run in a Java
  server, completely separated by the server running Django, an issue
  arises as how to communicate certificate/proxy location and
  passphrases back and forth from the Django decorator to the Java
  servlets.} however this poses a mild security threat: since the
servlets URLs must be public (so that the users' web browsers can
visit them), then an \ac{HTTP} request could be crafted to make the
servlets read/write the certificate/proxy file in an arbitrary
location on the filesystem.  Security is enforced with the following
procedure:
\begin{itemize}
\item Before starting the redirect to the a servlet, the Django access
  decorator creates an empty directory $L$, creates a ``marker''
  file in it, and writes a random string $K$ into this ``marker'' file.
\item The decorator redirects the web browser to the servlet \ac{URL},
  passing along $L$ and $K$ (as \ac{HTTP} cookies).
\item The servlet verifies that the ``marker'' file exists in $L$ and
  that it has the expected content $K$, then it deletes the ``marker''
  file and proceeds.  For added security, it can optionally verify
  that $L$ is a filesystem path starting with a configured prefix
  (e.g., \texttt{/var/www/portal}), so that possible damage is
  confined to a portion of the filesystem.
\end{itemize}
It is clear that the above procedure guarantees that hypothetical
attackers can only trick the \emph{GridCertLib} servlets into writing
into a location $L$ if and only if they can already write to $L$.

The added security layer is basically the only difference between the
Django-support servlets and the \emph{GridCertLib} example servlets
(see Section~\ref{example-servlets}).  After successful creation of
the certificate or proxy, the servlet redirects the web browser back
to the initial requesting page with no output.

As in the P-GRADE integration, two issues remain that might need
special attention in the future:
\begin{itemize}
\item The \ac{VOMS} configuration is the same for all users: while it is
  possible to extend the Django user object model to include
  individual \ac{VOMS} information, this is not necessary at present since
  all users of the \acs{GAMESS} portal belong to the same Virtual Organization.
\item Until the delegation feature becomes a standard feature of
  SWITCHaai, users have to select the special home organization \ac{VHO} in
  order to use the portal.
\end{itemize}

Django support for \emph{GridCertLib} provides an example of how
\emph{GridCertLib} can be integrated into an existing web framework
with little coding and only small edits to tune the example servlet
behavior to the interface expected by other portal components.


\section{Conclusions and Future Developments%
  \label{conclusions-and-future-developments}%
}

\emph{GridCertLib} is an easy to use Java library that enables
automatic creation of \ac{SLCS} certificates and/or Grid proxies from
\ac{SAML2} assertions obtained from successful Shibboleth
authentication. It can be integrated into real-world Grid portals,
hiding the complexities of X.509 certificate usage from the portal
user. This considerably lowers the barrier to Grid usage, potentially
allowing much larger communities to profit from Grid resources
securely.  Source code for \emph{GridCertLib} is publicly available
from
\href{http://gridcertlib.googlecode.com/}{\url{http://gridcertlib.googlecode.com/}}
under the Apache License version~2.0 \cite{ApacheLicence2}.

The current implementation of \emph{GridCertLib} relies on three key
features of the SWITCHaai infrastructure: Shibboleth authentication,
\ac{ID-WSF} \ac{ECP} delegation, and the \ac{SLCS} online \ac{CA}
service.  The integration of these three components together with a
valid access to a \ac{VOMS} server, allow the creation of any
community-specific web portal that can leverage the national grid
computing infrastructure \ac{SMSCG} \cite{SMSCG} thus enabling Grid
use by virtually any Swiss scientific community. 

An interesting future development could be to adapt GridCertLib to
draw certificates from the (recently created) \acsu{TERENA} on-line
\ac{CA}; this would lift the dependency on the Swiss infrastructure
and potentially allow usage of GridCertLib on any European Grid
infrastructure.  

More generally, one could investigate whether \emph{GridCertLib} could
be ported to provide its functionality on top of equivalent base
technologies (e.g., substitute Shibboleth with a different
\acs{SAML}-based federated authentication infrastructure).
Developments in this area could turn \emph{GridCertLib} into a modular
system capable of providing its functionality for almost all Grid
users today.  No investigation has been carried out by us in
this area: the project that funded \emph{GridCertLib} development had
a practical scope of producing a simple single sign-on solution for
the selected portals; we are anyway open to collaborations in this
respect. 

\emph{GridCertLib} has already been successfully deployed and
integrated into a Bioinformatics portal based on P-GRADE, and into a
Django-based Computational Chemistry portal, proving the flexibility
and re-usability of the library and its design.

We will assist in the integration of \emph{GridCertLib} into portals that
are in use in Switzerland, like JOpera \cite{JOpera} and the new WS-PGRADE
\cite{WsPgrade}. We will consider requests for extensions in functionality
of the \emph{GridCertLib} based on the experience with these new portals.

Looking further into the future, \emph{GridCertLib} will greatly
profit from the upgrade of the SWITCHaai federation to the next
version of Shibboleth: this will enable true single-sign on and Grid
usage in one portal, without the need to use a special \ac{VHO}
account.  The SystemsX project SyBIT \cite{SyBIT} also plans to
upgrade its P-GRADE portal from the current Gridsphere-based
implementation to the more modern WS-PGRADE, which makes use of the
Liferay portal \cite{Liferay} technology: besides many portal-related
improvements, this will allow the users to freely choose the \ac{VOMS}
attributes they wish to associate with their proxy.  However, due to
the entirely new portal code base, a new programming effort will be
needed to integrate \emph{GridCertLib} into the Liferay framework.


\subsection*{Acknowledgements}

This work was carried out in the context of the ``Swiss Grid Portal''
project, funded through the SWITCH-AAA track and through the SyBIT
project of SystemsX.ch.  We would like to thank all our collaborators
in the Swiss Grid Portal project ---Cesare Pautasso, Fr\'ed\'erique
Lisacek, Heinz Stockinger--- and also all the help we received from
the Hungarian Academy of Sciences SZTAKI for the integration with
P-GRADE, especially \'Akos Balasko.


\section*{List of acronyms%
  \phantomsection%
  \addcontentsline{toc}{section}{List of acronyms}%
  \label{acronyms}%
}


\begin{acronym}[MMMMMMM]
  \acro{AAI}{Authentication and Authorization Infrastructure}%
  \acro{AC}{Attribute Certificate \acroextra{(VOMS, X.509)}}%
  \acro{API}{Application Programming Interface}%
  \acro{CA}{Certification Authority}%
  \acro{CSR}{Certificate Signing Request}%
  \acro{DN}{Distinguished Name}%
  \acro{ECP}{Enhanced Client or Proxy}%
  \acro{EGEE}{Enabling Grids for E-sciencE}%
  \acro{FQAN}{Fully-Qualified Attribute Name \acroextra{(VOMS)}}%
  \acro{GAMESS}{General Atomic and Molecular Electronic Structure System, a Computational Chemistry application (see \cite{GAMESS:1993,GAMESS:2005})}%
  \acro{GC3}{Grid Computing Competence Center, University of Zurich}%
  \acro{HTTP}{HyperText Transfer Protocol}%
  \acro{ID-WSF}{Identity Domain - Web Service Framework \acroextra{(Shibboleth)}}
  \acro{IGTF}{International Grid Trust Federation}%
  \acro{IdP}{Identity Provider \acroextra{(Shibboleth)}}%
  \acro{MAMS}{Meta Access Management System, an Australian development project \acroextra{(see \cite{MAMS})}}%
  \acro{PKI}{Private Key Infrastructure}%
  \acro{SAML}{Security Assertion Markup Language}%
  \acro{SAML2}{Security Assertion Markup Language version 2}%
  \acro{SARoNGS}{UK development project (see \cite{SARoNGS})}%
  \acro{SLCS}{Short-Lived Credential Service}%
  \acro{SMSCG}{Swiss Multi-Science Computing Grid}%
  \acro{SP}{Service Provider \acroextra{(Shibboleth)}}%
  \acro{SSO}{Single Sign-On}%
  \acro{SWITCH}{Swiss Academic Network Provider}%
  \acro{TERENA}{Trans-European Research and Education Networking Association}%
  \acro{URL}{Uniform Resource Locator}%
  \acro{VHO}{Virtual Home Organisation}%
  \acro{VOMS}{Virtual Organisation Membership Service}%
  \acro{VO}{Virtual Organisation}%
  \acro{XACML}{eXtensible Access Control Markup Language}%
\end{acronym}




\end{document}